\documentclass[11pt]{JHEP3}
\usepackage[pdftex]{graphicx}
\usepackage{caption}
\usepackage{amsmath}
\usepackage{amssymb}

\title{Radiative Fermion Masses in Local D-Brane Models }
\author{C.P.~Burgess$^{1,2}$, Sven Krippendorf$^{3}$, Anshuman Maharana$^{3}$, and Fernando Quevedo$^{3,4}$\\
$^{1}$ Department of Physics \& Astronomy, McMaster University, Hamilton ON, Canada.\\
$^{2}$ Perimeter Institute for Theoretical Physics, Waterloo ON, Canada.\\
$^{3}$ DAMTP, University of Cambridge, Wilberforce Road, Cambridge, CB3 0WA, UK.\\
$^{4}$ ICTP, Strada Costiera 11, Trieste 34151, Italy.}
\preprint{DAMTP-2011-10\\
}
\abstract{In the context of D-brane model building, we present a realistic framework for generating fermion masses that are forbidden by global symmetries. We show that the string theoretical Large volume scenario circumvents the standard lore that fermion masses generated by loop effects are too small in generic gravity mediated  scenarios. We argue that the fact that in toric singularity models, the up quark masses have always a zero eigenvalue, corresponding to the lightest generation, is due to the presence of approximate global symmetries that we explicitly identify in del Pezzo singularities. These symmetries are broken by global effects and therefore proportional to inverse powers of the volume. We estimate the generic size of radiative corrections to fermion masses in different phenomenological manifestations of the Large volume scenario. Concrete realizations in terms of flavor violating soft-terms are estimated and contrasted with current bounds on flavour changing neutral currents. Contributions from  generic extra Higgs-like fields set bounds on their masses close to the GUT scale to produce realistic fermion masses.}


\begin{document}

\section{Introduction}

Branes at singularities provide a promising avenue for constructing models of particle physics from string theory. In the framework of IIB string theory they have been used to obtain supersymmetric extensions of the Standard Model and various GUT models~\cite{0005067, 0105042, 0312051, 0508089, 0703047, 0810.5660, 0910.3616}. Recent progress has also been made with these models in obtaining hierarchical fermion masses and realistic hierarchies in the CKM matrix~\cite{1002.1790}. (Fermion mass hierarchies have also been studied in intersecting brane models~\cite{Kiritsis,  Cvetic:2009ez} and in F-theory~\cite{Cecotti:2009zf}.)

The models constructed in~\cite{1002.1790} all predict that the mass of the lightest up-type quark vanishes at tree level. In this paper we trace this prediction to the existence of an approximate $U(1)$ symmetry of the relevant superpotential, which forbids the $u$-quark mass term. This symmetry is a global symmetry of the local model, which is expected to be broken once the model is embedded into a compact bulk~\cite{Banks:1988yz, 0805.4037, Bankst}. Consequently, loop effects involving bulk fields should generically generate a mass term, raising the potential that light fermion masses might be radiatively generated in these models.

We estimate the size of these radiative corrections in the context of the Large volume compactifications (LVS)~\cite{0502058}, for which the various moduli are calculably stabilised at large values of the extra-dimensional volume. These compactifications naturally break supersymmetry, which is fed through to the observed sector through the interactions of the moduli (a form of gravity-modulus mediated SUSY breaking~\cite{0610129,0906.3297}). In particular, knowledge of the volume-dependence of the various moduli masses allows us to estimate the volume-dependence that suppresses the size of the loop-generated fermion masses.

What is interesting is that our estimates differ from estimates of radiative corrections to light fermion masses that were obtained some time ago~\cite{Ibanez:1982xg} subject to a set of then-reasonable assumptions about a ``generic gravity mediated scenario". In these earlier `generic' estimates the size of the fermion masses generated turns out to be too small to be relevant for phenomenology; an observation that has come to be regarded as a `no-go' result for obtaining fermion masses radiatively in models with gravity-mediated supersymmetry breaking.

Various scenarios for SUSY breaking~\cite{0610129, 0906.3297,0505076,cjp, Choi} have been discussed within the LVS context. While the basic mechanism for SUSY breaking is that of gravity mediation, the structure of soft masses found is much richer (due to the approximate no-scale structure in IIB string theory) than is assumed by the old generic arguments. A consequence is that loop corrections to fermion masses can be significantly larger than the estimate in~\cite{Ibanez:1982xg}.

The rest of the paper is organised as follows. We start with a brief review of models from branes at toric singularities and their effective field theory once embedded in a global setup. Next, we examine the Yukawa matrix in explicit models constructed on del Pezzo singularities and isolate the global $U(1)$ symmetry of the local model associated with the zero eigenvalue of the mass matrix. We then discuss bulk effects which break this symmetry. Turning to the discussion of radiatively generated fermion masses, we next review the old `no-go' results that arose within the context of GUT model building~\cite{Ibanez:1982xg}. We then re-examine these arguments in the context of the LVS and show that the mass generated is significantly higher than for generic gravity-mediated scenarios, although obtaining a realistic value requires meeting several challenges, not least of which is the issue of flavour-changing neutral currents (FCNCs), on which we elaborate in more detail below.

\section{Branes at singularities}

The past decade in string model building has seen enormous developments in local modular string model building, initiated in~\cite{0005067}. In this approach the Standard Model degrees of freedom arise from open strings associated with D-branes localised in a certain region of the internal manifold. The main advantage of these models is that properties of the gauge theory such as the gauge coupling are only sensitive to local properties of the compactification geometry and insensitive to the whole bulk which is notoriously hard to control. 

For example the strength of the Standard Model gauge couplings can be obtained from the size of the cycles the branes are wrapping and are independent of the overall bulk volume. In other words, while for example the overall bulk volume can go to infinity the gauge couplings are only sensitive to the local geometry and remain finite. Within this philosophy very promising models have been developed with branes at singularities in type IIB string theory~\cite{0005067, 0508089, 0810.5660, 1002.1790} and more recently in F-theory (for a review see~\cite{0911.3008, 1001.0577}). We focus on the former from now on.

There is a great understanding of the low-energy gauge theory arising from branes at singularities in type IIB string theory. The spectrum can be thought of as arising from open string states stretching between the branes. Before giving concrete examples, let us summarise the general ingredients and tools. Two types of branes, so-called fractional $D3$ and fractional $D7$ branes, give rise to the gauge theory. The former are bound states of $D3,$ $D5$ (wrapping collapsed 2-cycles) and $D7$ branes wrapped on the collapsed 4-cycle. The latter wrap both the collapsed cycle and a bulk cycle. In the absence of O-planes passing through the singularity, the gauge groups are unitary gauge groups $U(N).$ The matter content for such brane models comes in chiral superfields transforming in bi-fundamentals and vector superfields transforming in the adjoint. 

One can summarise the matter content of the theory using a quiver diagram. There are two types of chiral superfield states at the singularity, one arising from strings starting and ending on $D3$ branes and the other arising from strings either starting on $D3$ branes and ending on $D7$ branes or vice versa. The matter content and superpotential can most conveniently be obtained with dimer diagrams that relate to the topological properties of the singularity, see~\cite{0803.4474, 1002.1790} for a more detailed discussion.

It turns out that isometries or symmetries of the singularities are present in the superpotential of the gauge theory as global symmetries.\footnote{Although global symmetries are possible for non-compact constructions -- for which Newton's constant is effectively set to zero -- for a full string construction, including compactification of the extra dimensions, they are either only approximate symmetries, or gauged~\cite{Banks:1988yz, 0805.4037}. We return to the consequences of this observations below.} These global symmetries have interesting phenomenological consequences for model building. In section~\ref{globalsym} we shall find that the zero mass for the lightest generation at tree-level is a result of one of these global symmetries in the superpotential. To illustrate how global symmetries restrict the superpotential for del Pezzo singularities, recall that the $n$-th del Pezzo surface $dP_n$ corresponds to $\mathbb{P}^2$ blown-up at $n$ generic points. Considering the zeroth del Pezzo surface $dP_0=\mathbb{C}^3/\mathbb{Z}_3$, the gauge group is a product of three $U(N)$ factors and the matter content is three copies of bi-fundamentals under each pair of gauge groups $X_{i}, Y_{i}, Z_{i} $ with $(i = 1,2,3)$. Gauge invariance requires the cubic part of the superpotential to be of the form
\begin{equation}
W=c_{ijk}X_i Y_j Z_k+W_{\rm D3D7}
\end{equation}
where $c_{ijk}$ is arbitrary. However $dP_0$ possesses an $SU(3)$ isometry, which in turn requires $c_{ijk}=\epsilon_{ijk}$. This is broken to $SU(2)\times U(1)$ in $dP_1$ and further to $U(1)$ in $dP_2.$ There are no such isometries in higher del Pezzo surfaces, but, as studied in~\cite{0404065}, there are new hidden global symmetries, more precisely an $E_n$ global symmetry for a $dP_n$ singularity and they together with gauge invariance determine the superpotential completely. To complete this survey of the setup, let us discuss a low-energy model on the first del Pezzo surface $dP_1.$

\subsection{Standard model on $dP_1$}
\label{dp1}

Phenomenological string constructions built using $dP_1$ were discussed in refs.~\cite{0810.5660, 1002.1790}. Choosing (for now) the Standard Model gauge interactions to be included within the gauge group $U(3)\times U(2)\times U(1) \times U(1)$ in $dP_1$ we can write the superpotential as follows:
\begin{eqnarray}
 \nonumber W_{dP_1} &=& X_{23} Y_{31} Z_{12} - X_{12} Y_{31} Z_{23}
 + X_{36} Y_{62} Z_{23} - X_{23} Y_{62} Z_{36}\\ \nonumber 
 && -X_{36} Y_{23} Z_{12} \frac{\Phi _{61}}{\Lambda} 
 + X_{12} Y_{23} Z_{36} \frac{\Phi _{61}}{\Lambda}+W_{\rm D3D7}\\
 &=&\left( \begin{array}{c}
 X_{23} \\
 Y_{23} \\
 Z_{23}
 \end{array} \right)
 \left( \begin{array}{ccc}
 0 & Z_{12} & -Y_{62} \\
 -Z_{12} \frac{\Phi _{61}}{\Lambda} & 0 & X_{12} \frac{\Phi _{61}}{\Lambda} \\
 Y_{62} & -X_{12} & 0
\end{array} \right)
\left( \begin{array}{c}
 X_{36} \\
 Y_{31} \\
 Z_{36}
\end{array} \right) +W_{\rm D3D7} \,.
\label{dp1super}
\end{eqnarray}
where $(X_{23},Y_{23},Z_{23})$ are left-handed quarks, $(X_{36}, Y_{31}, Z_{36})$ are right-handed quarks and the matrix $Y$ is the Yukawa matrix.

\begin{center}
\includegraphics[width=0.55\textwidth]{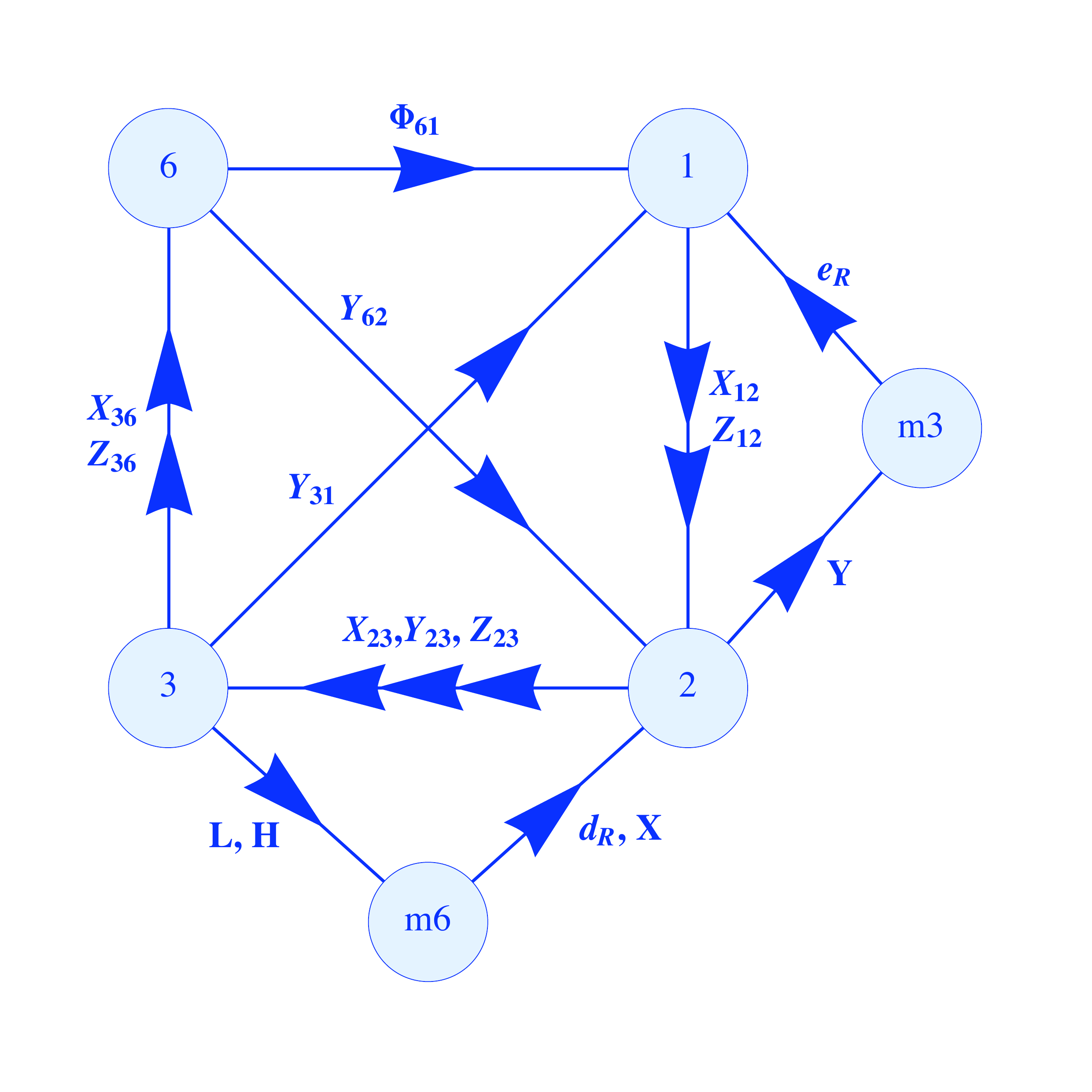}
\captionof{figure}{\footnotesize{The $dP_1$ quiver with two additional $D7$ gauge groups. An arrow between any given nodes corresponds to bi-fundamental matter. An arrow starting at node $i$ and ending at node $j$ corresponds to a field $X_{ij}$ transforming as $(N_j,\bar{N}_i).$ Identifying gauge groups $1$ and $6$ with $U(1)$ symmetries, gauge group $3$ with $U(3)$ and gauge group $2$ with $U(2)$ leads to Standard Model matter content as discussed in~\cite{0810.5660}.}\label{fig:dp1quiver}}
\end{center}
Given this superpotential, one can analyse the mass and mixing structure arising from this non-trivial Yukawa matrix~\cite{0810.5660, 1002.1790}. It was found in~\cite{1002.1790} that a superpotential of the form of (\ref{dp1super}) has enough structure to allow for both hierarchical mass structure and mixing in the quark sector. The quark mass matrix has eigenvalues $(M,m,0)$ with $M \gg m$, the hierarchy  given by the parameter $\Phi / \Lambda$.
However, we generally encounter a zero eigenvalue for the lightest generation of $D3-D3$ up-quarks, which will relate to global symmetries of the gauge theory later in section~\ref{globalsym}.

\section{The effective field theory of local models in the Large Volume}

To construct a realistic model one has to embed local models into a compactification where all moduli are stabilised. For this purpose we use the Large volume model (LVS)~\cite{0502058} as a convenient framework for stabilisation.

\subsection{The Large volume scenario}

The LVS is a mechanism for modulus stabilisation in type IIB Calabi-Yau flux compactifications. In generic type IIB flux compactifications, the presence of background 3-form fluxes on their own can stabilise the dilaton and the complex structure moduli of the underlying Calabi-Yau geometry~\cite{0105097}. In a second step the K\"ahler moduli can be stabilised by non-perturbative effects localised on four cycles associated with various branes that source the geometries~\cite{kklt, 0502058}. The LVS identifies an interesting subclass of these stabilisations for which the volume modulus ${\cal V}$ is naturally stabilised at a volume that is exponentially sensitive to the size of a smaller blow-up cycle, $\tau_s$,
\begin{equation}
 {\cal V}\sim {\rm exp}(a \tau_s) \,.
\end{equation}
Here $\tau_s$ is the size of a blow-up cycle for a point-like singularity in the underlying geometry, whilst $a$ is an order-unity constant.

A key ingredient is the inclusion of the leading order $\alpha'$ correction to the 4D K\"ahler potential and non-perturbative contributions to the superpotential, since this is what generates the potential including a minimum with compactification volumes that are exponentially large in $\tau_s$. $\tau_s$ itself scales as the inverse of the value of the dilaton
\begin{equation}
 \tau_s\sim\frac{1}{g_s} \,,
\end{equation}
and hence is given as a ratio of integer flux quanta. Thus the framework naturally generates an exponentially large value in the volume modulus of the compactification from integer flux quanta, with ${\cal V}$ passing through an exponentially large range of values as the fluxes run through a more moderate range.

One can show on general grounds~\cite{0805.1029} that a minimum for the K\"ahler moduli exists at large volume if there is (i) at least one K\"ahler modulus as blow-up mode, and (ii) the number of complex structure moduli is greater than the number of K\"ahler moduli. The Large volume scenario hence allows for local models placed at blow-up modes. The brane physics on these blow-up modes decouples at leading order from physics arising from other blow-up cycles.
Scenarios in which the Standard Model degrees of freedom are localised at a singularity require the modulus corresponding to the blow up cycle to be stabilised at zero size, which is favoured by the structure of D-terms, the associated Fayet-Iliopoulos parameter is the size of the blow-up cycle.

\subsection{The effective field theory for local models in Large volume compactifications}
We consider two different possibilities with regard to the realisation of the Standard Model, following~\cite{0906.3297}. In the first the Standard Model degrees of freedom arise from branes localized at singularities. In this case the modulus corresponding to blowing up the singularity is in the sub-stringy regime. The effective field theory
is written as an expansion around the zero value of the modulus. In the second case the Standard Model brane wraps a modulus which is in the geometric regime. In this case the effective field theory is valid for values of this modulus hierarchically larger than the string scale. The low energy superpotential, K\"ahler potential and gauge kinetic functions have been described in~\cite{0906.3297}.

Our interest is in radiative corrections (for fermion masses) generated via SUSY breaking. Thus the structure of the F-terms and the coupling between the Standard Model sector and the  SUSY breaking sector shall be relevant. On this, the different realisations of the Standard Model (singular or geometric regime) do not affect the structure of F-terms but there can be two different scenarios depending on the location of the Standard Model construction in the bulk:
\begin{itemize}
\item{} {\it Scenario A:} if the dominant F-term is of the modulus on
which the Standard Model is localized, then the coupling between the Standard Model sector and the SUSY breaking sector is suppressed by the string scale ($M_{\rm UV} = M_{\rm string}$);
\item{} {\it Scenario B:} on the other hand if the dominant F-term is of a modulus which is geometrically separated in the extra dimensions from the Standard Model cycle, the suppression is  by the Planck scale ($M_{\rm UV} = M_{\rm Planck}$).
\end{itemize}
The explicit form of the F-terms and pattern of soft masses can be found in~\cite{0610129, 0906.3297}.

\section{Global symmetries and zero masses}
\label{globalsym}

A generic feature of models constructed from branes at singularities is the presence of global symmetries in the superpotential. These symmetries can arise due to symmetries of the local geometry or anomalous $U(1)$ symmetries acquiring masses via the St\"uckelberg mechanism. For example in the case of branes probing del Pezzo singularities, the following symmetries are present:
\begin{itemize}
\item For the first del Pezzo surfaces there are global isometries of the singularity, whose symmetry decreases with increasing number of points blown-up on $\mathbb{P}^2.$ In $dP_0,$ there is an $SU(3)$ isometry, which reduces in $dP_1$ to $SU(2)\times U(1).$ In $dP_2$ only one $U(1)$ symmetry is present. Any higher del Pezzo does not possess such isometries.
\item At the high scale (e.g. string scale) any anomalous $U(1)$ factor of a $U(N)$ gauge group obtains a string scale mass via the standard Green-Schwarz mechanism and in the low-energy theory these symmetries will be present as global symmetries.
\item In~\cite{0404065} it was shown that for a $dP_n$ singularity there is an associated $E_n$ global symmetry and a $U(1)_R$ symmetry.
\end{itemize}
These symmetries can have important phenomenological implications. For instance, it has been argued that they can be relevant for the stability of the proton~\cite{9908305}. Below we shall find that the zero mass for the lightest quark generation in the del Pezzo models is a consequence of these global symmetries. Focusing on the model on the first del Pezzo surface $dP_1$ described in section~\ref{dp1}, we  show that global symmetries forbid the mass for the up-quark. We can do this by going from the gauge eigenbasis with superpotential as in equation~\ref{dp1super} to the mass eigenbasis. A similar analysis can be carried out for all higher toric del Pezzo singularities but becomes more technical and is not illuminating for the reader. We discuss the example of the second del Pezzo surface in Appendix A.

We begin by diagonalising the Yukawa matrix  (\ref{dp1super}) by a bi-unitary transformation $A^T Y B=:D.$ To leave the superpotential invariant the left-handed quarks transform as $Q_L^m\to A^* Q_L$ and the right handed quarks as $q_R^m=B^\dagger q_R.$ To find the matrices $A$ and $B$ it is useful to consider
\begin{eqnarray}
 D D^\dagger&=&(A^T YB)(B^\dagger Y^\dagger A^*)=A^T YY^\dagger A^*\; \text{ and}\\
 D^\dagger D&=&(B^\dagger Y^\dagger A^*) (A^T YB)=B^\dagger Y^\dagger Y B\, .
\end{eqnarray}
For $dP_1$ we find $A$ and $B$ to be given by
\begin{eqnarray}
 A&=&\left(
 \begin{array}{ccc}
 a_1 X_{12} \Phi _{61} & -a_2 Z^\dagger_{12} & a_3 Y^\dagger_{62} X_{12} \\
 a_1 Y_{62} & 0 & -a_3X^\dagger_{12} \Phi^\dagger_{61} X_{12}-Z^\dagger_{12} \Phi^\dagger_{61} Z_{12} \\
 a_1 Z_{12} \Phi _{61} & a_2 X^\dagger_{12}& a_3Y^\dagger_{62} Z_{12}
 \end{array}
 \right),\\
 B&=&\left(
 \begin{array}{ccc}
 b_1 X_{12} & b_2Y^\dagger_{62} X_{12} & -b_3Z^\dagger_{12} \\
 b_1 Y_{62} & -b_2X^\dagger_{12} X_{12}-b_3Z^\dagger_{12} Z_{12} & 0 \\
 b_1 Z_{12} & b_2Y^\dagger_{62} Z_{12} & b_3X^\dagger_{12}
 \end{array}
 \right),
\end{eqnarray}
where $a_i$ and $b_i$ denote the normalisation prefactor of each column vector. In this basis the superpotential, up to normalisation factors becomes
\begin{eqnarray}
 \nonumber W&=&\left(
 \begin{array}{c}
 X^\dagger_{12} \frac{\Phi^\dagger_{61}}{\Lambda} X_{23}+Y^\dagger_{62} Y_{23}+Z^\dagger_{12} \frac{\Phi^\dagger_{61}}{\Lambda} Z_{23} \\
 -X_{23} Z_{12}+X_{12} Z_{23} \\
 X^\dagger_{12} X_{23} Y_{62}+Z^\dagger_{12} Y_{62} Z_{23}-Y_{23}\frac{\Phi _{61}}{\Lambda} \left(|X_{12}|^2+|Z_{12}|^2\right)
 \end{array}
 \right).\\ \nonumber &&\left(
 \begin{array}{ccc}
 0 & 0 & 0 \\
 0 & \sqrt{|X_{12}|^2+|Y_{62}|^2+|Z_{12}|^2} & 0 \\
 0 & 0 & -\sqrt{|Y_{62}|^2+\left(|X_{12}|^2+|Z_{12}|^2\right) \frac{|\Phi _{61}|^2}{\Lambda^2}}
 \end{array}
 \right).\\ &&\left(
 \begin{array}{c}
 X^\dagger_{12} X_{36}+Y^\dagger_{62} Y_{31}+Z^\dagger_{12} Z_{36} \\
 X^\dagger_{12} X_{36} Y_{62}-Y_{31} \left(|X_{12}|^2+|Z_{12}|^2\right)+Z^\dagger_{12} Y_{62} Z_{36} \\
 -X_{36} Z_{12}+X_{12} Z_{36}
 \end{array}
 \right)
\end{eqnarray}
Now we can see that a potential mass term for the lightest generation is not allowed by gauge symmetry and the $U(1)$ flavour symmetry. This can been seen from  Table~\ref{tab:gaugecharges} which summarises all gauge charges of the original fields and the mass eigenstates of the quarks.
\begin{center}
\begin{tabular}{c|ccccc}
 Field & $n_1$ & $n_2$ & $n_3$ & $n_6$ & Flavour $U(1)$ \\ \hline \hline
 $X_{23}$ & 0 & 1 & -1 & 0 & 0 \\
 $ Y_{23}$ & 0 & 1 & -1 & 0 & $\theta$ \\
 $ Z_{23}$ & 0 & 1 & -1 & 0 & 0 \\ \hline
 $X_{36} $& 0 & 0 & 1 & -1 & 0 \\
 $ Y_{31} $& -1 & 0 & 1 & 0 & 0 \\
 $ Z_{36} $& 0 & 0 & 1 & -1 & 0 \\ \hline
 $ X_{12} $& 1 & -1 & 0 & 0 & 0 \\
 $ Y_{62} $& 0 & -1 & 0 & 1 & 0 \\
 $ Z_{12} $& 1 & -1 & 0 & 0 & 0 \\ \hline
 $\Phi_{61}$& -1 & 0 & 0 & 1 & $-\theta$\\ \hline
 $ Q_{r1}^m$ & -1 & 1 & 1 & -1 & 0 \\
 $ Q_{r2}^m$ & -1 & 0 & 1 & 0 & 0 \\
 $ Q_{r3}^m$ & 1 & -1 & 1 & -1 & 0 \\ \hline
 $Q_{L1}^m$& 0 & 2 & -1 & -1 & 0 \\
 $ Q_{L2}^m$& 1 & 0 & -1 & 0 & 0 \\
 $ Q_{L3}^m $& -1 & 1 & -1 & 1 & 0
\end{tabular}
\captionof{table}{\footnotesize{Summary of the $U(1)$ charges of each field in $dP_1.$ The last six are the charges for the quarks in the mass eigenstate basis.}\label{tab:gaugecharges}}
\end{center}

As mentioned earlier, we know that theories of quantum gravity cannot possess any exact global symmetries. In the context of string theory this was first pointed out in~\cite{Banks:1988yz} and recently extended to models including D-branes in~\cite{0805.4037}. Thus the global symmetries discussed above are approximate and only present as global symmetries in the low-energy effective action. They are broken by bulk effects (recall that the infinite volume limit of a compactification corresponds to taking Newton's constant to infinity, hence the breaking of the global symmetries is not seen in the purely local analysis). In the case of isometries this manifests itself in a difference in properties of compact and non-compact Calabi-Yaus (with $SU(3)$ holonomy): non-compact Calabi-Yaus can possess isometries (e.g. $\mathbb{C}^3/\mathbb{Z}_3$ has an $SU(3)$ symmetry), however compact Calabi-Yaus do not possess any isometries.

One can consider non-compact Calabi-Yaus with isometries and embed them in a compact Calabi-Yau. In this case the isometries are broken by compactification effects. By taking a large volume limit one can obtain a metric where the isometries are present as approximate symmetries of the local geometry and they become exact in the limit of infinite volume (corresponding to infinite Newton's constant and hence decoupled gravity). By considering local models in such a setting, at leading order (in an inverse volume expansion) one has global symmetries. At higher order, one expects these symmetries to be broken via couplings to the bulk, such effects shall play a central role in generating the non-zero masses for the lightest generation.

\section{Flavour Changing Effects}

At tree-level heavy generations have a mass and we are interested in radiative corrections which mediate the mass to the lightest generation; thus we need to consider processes which change flavour. The purpose of this section is to identify flavour changing effects in our models and discuss experimental constraints on them. Here there are two possible contributions for flavour change: one arising from potential entries in the sfermion-mass matrix that are not proportional to the fermion mass matrix, the other arising from additional Higgs-doublets which are generic for models from branes at singularities~\cite{0005067,1002.1790}.

\subsection{Flavour Change from sfermion masses}

A generic flavour-blind choice of sfermion masses can lead to unobserved large flavour changing processes~\cite{Hall:1985dx}. A viable scenario of supersymmetry breaking should suppress these flavour changing processes significantly (cf. Figure~\ref{fig:mesongraphs}). In the context of sfermion masses in gravity mediated supersymmetry breaking this can be achieved if the sfermion mass matrix $m_{\alpha\beta}$ is proportional to the K\"ahler metric of these matter fields $K_{\alpha\beta}.$ This restricts the structure of allowed gravity mediated scenarios severely, but one can show that the effective action of type IIB string compactifications and, in particular, in LVS naturally hints to a solution of this supersymmetric flavour problem~\cite{0710.0873}. This is achieved since supersymmetry is only broken by the K\"ahler moduli whereas the flavour structure is generated by complex structure moduli.

For example taking a Calabi-Yau with one K\"ahler modulus and a matter field localised on a $D3$ brane, the effective supergravity setup is given by
\begin{equation}
 K=-3\log{(T+\bar{T})}-i\log{\left(\int\Omega\wedge\bar{\Omega}\right)}
 -\log{(S+\bar{S})}+\frac{1}{(T+\bar{T})}(1+f(U,\bar{U}))_{\alpha\beta}
 \Phi_\alpha\bar{\Phi}_\beta\, ,
\end{equation}
where in the matter metric there is an overall factor depending on the K\"ahler moduli and an individual piece depending on the complex structure moduli $U$ which can be flavour off-diagonal. A K\"ahler potential of the above form has a sfermion mass matrix which is proportional to the matter metric. As highlighted in~\cite{0912.2950}, this may not be the case for compactifications with several moduli such as swiss-cheese LVS with two K\"ahler moduli where the K\"ahler potential takes the following form
\begin{equation}
 K=-2\log{\left[(\tau_b-\omega^b_{\alpha\beta} \bar{\phi}_\alpha\phi_\beta)^{3/2}-(\tau_s-\omega^s_{\alpha\beta} \bar{\phi}_\alpha\phi_\beta)^{3/2}\right]}\, ,
\end{equation}
where $\omega_b$ and $\omega_s$ are arbitrary functions. This could lead to flavour changing effects but their presence and size depends on the exact structure of $\omega_b$ and $\omega_s$ which is not known. Besides the corrections to the K\"ahler potential there can be similar effects for example arising from non-perturbative corrections to the superpotential which lead to flavour violating effects via A-terms~\cite{0312157, 1012.1858}. We will work with the assumptions of~\cite{0912.2950} that such effects exist but are suppressed in terms of the overall volume.

\subsection{Flavour Change from additional Higgses}

Recall that in the Standard Model the diagonalisation of the quark mass matrix also brings the interactions between the Higgs and the quarks to a diagonal form. As a result there are no tree-level flavour changing processes mediated by the Higgs. This is no longer true in theories with multiple Higgs fields. The discussion of flavour changing effects due to multiple Higgs fields is best done in the so-called Runge basis~\cite{0909.4541,0912.0267}. In this basis the Higgses are parametrised by $\varphi_{\rm vev}$ (the linear combination of Higgs fields that acquires a vev) and $\varphi_{\perp}^i$ (linear combinations of the remaining Higgs fields which do not acquire a vev and are taken to be orthogonal to $\varphi_{\rm vev}$ and each other in field space). While the field corresponding to the fluctuations of $\varphi_{\rm vev}$ does not induce any flavour change, the couplings of $\varphi_{\perp}^i$ are non diagonal in flavour space. 

In particular, the neutral Higgses associated with $\varphi_{\perp}^i$ have couplings of the form
\begin{equation}
 \tilde{Y}^i_{ab}\varphi_{\perp}^i Q_L^a q_r^b\, ,
\end{equation}
where $a, \, b$ are generation indices. The strongest bounds on flavour changing processes arise from the mixing of neutral mesons: As shown in Figure~\ref{fig:mesongraphs} a tree-level flavour changing neutral current contribution from the orthogonal Higgs competes with the Standard Model 1-loop box diagrams.
\begin{center}
\begin{tabular}{c c c}
\includegraphics[width=0.3\textwidth]{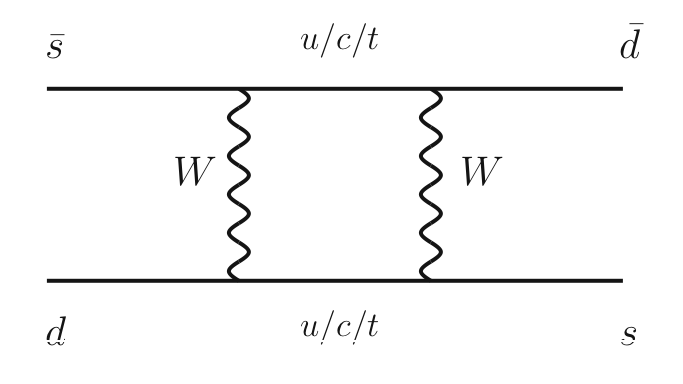} & \includegraphics[width=0.3\textwidth]{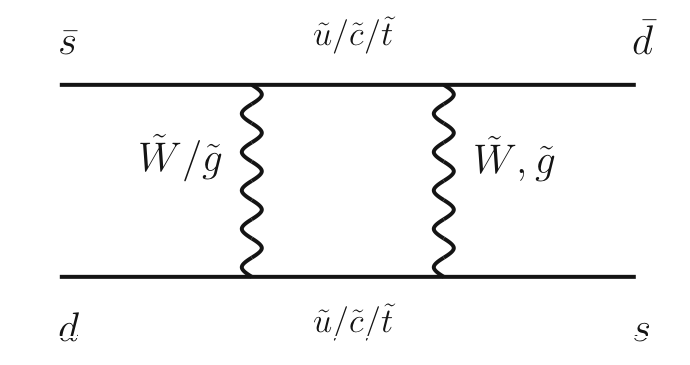} & \includegraphics[width=0.25\textwidth]{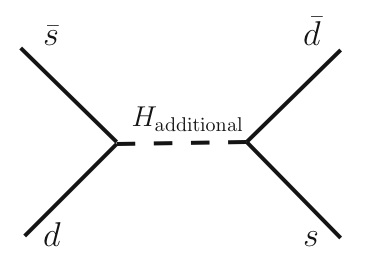}
\end{tabular}
\captionof{figure}{\footnotesize{ {\bf Left:} Flavour change in the Standard Model. {\bf Middle:} Supersymmetric version of diagram on the left. Off-diagonal contributions in the scalar mass matrix can lead to additional contributions involving gluinos. {\bf Right:} Contribution from the additional Higgs to the same FCNC process.}\label{fig:mesongraphs}}
\end{center}
Working in the Standard Model limit (with one light Higgs, corresponding to the Standard
Model Higgs and all other additional Higges being parametrically heavier) a generic analysis based on a classification of the higher dimensional operators~\cite{1002.0900} requires that the masses of the addtional Higgs are above $10^{3}$ TeV. 

There are two scenarios for  masses of ``additional'' Higgs fields:
\begin{enumerate}
\item The mass is generated by a $D3D3-D3D7-D7D3$ coupling, where one of the $D3D7$ states has a vev, hence the superpotential coupling $W\supset H_{\rm additional} A_{\rm D3D7}B_{\rm D7D3}$ becomes a mass term. Generically this vev is of order the string scale and hence induces a mass for the additional Higgs field of that order.
\item A $\mu$-term can be generated by euclidean $E3$ branes intersecting with the singularity~\cite{0711.1316}. The coupling is of the type $W\supset A H_u H_d e^{-a T_s}$ where the size of this contribution is determined by the size of the cycle the $E3$ brane is wrapping, parametrised by $T_s.$ In a LVS framework, this corresponds to some inverse power of the volume and hence while still large a hierarchically smaller contribution to the masses for the additional Higgses.
\end{enumerate}

\section{Fermion Masses from radiative corrections}
\label{estim}

Given that an approximate symmetry ensures the lightest generation is massless at tree level, it is natural to ask whether radiative corrections can be relevant for generating their masses. Recall that the relevant symmetry arose as a global symmetry intrinsic to the singularity. Interactions including bulk fields generically do not respect these symmetries and so effective terms in the Lagrangian breaking these symmetries can be generated.

We shall focus on effective terms associated with SUSY breaking. In this section we review the basic structure of such terms and discuss old arguments why such corrections to fermion masses have generically hierarchically smaller values than the experimentally observed values~\cite{Ibanez:1982xg}.

Let us begin by discussing how scalar masses are generated by radiative corrections in an effective supergravity setup. Scalar masses are completely determined by the supergravity scalar potential, which for the purpose of this article shall be assumed to be given by the F-term scalar potential with vanishing cosomological constant.  The general expression for scalar masses is
\begin{equation}
 m_{\alpha\beta}^2=m_{3/2}^2+V_0-F^m F^{\bar{n}}\partial_m\partial_{\bar{n}}\log{Z_{\alpha\beta}}\, , \label{scalarmass}
\end{equation}
where $Z_{\alpha\beta}$ is the moduli dependent prefactor of the canonical K\"ahler potential.

In typical setups this puts the scalar mass at the order of the gravitino mass and masses of the scalar component of a chiral superfield $C$ can be captured from an effective coupling of the following type:
\begin{equation}
 \frac{1}{M_{\rm UV}^2}CC^\dagger M M^\dagger\, ,
 \label{scalarmassterm}
\end{equation}
where $M$ denotes the field breaking supersymmetry. The D-term component of this expression then gives scalar masses of order the gravitino mass
\begin{equation}
 \left. \frac{1}{M_{\rm UV}^2}CC^\dagger M M^\dagger\right|_{\theta\theta \bar{\theta}\bar{\theta}} \sim \frac{1}{M_{\rm UV}^2} C C^\dagger F_M \bar{F}_M\sim m_{3/2}^2 C C^\dagger\, .
\end{equation}
We would like to emphasise that the above estimate assumes no cancellations between different contributions in equation~\ref{scalarmass}. Under special circumstances such as in (extended) no-scale models such cancellations appear, rendering scalars significantly lighter than the gravitino mass.

Now turning to fermion masses we directly see that the origin for radiative corrections has to be different from equation~\ref{scalarmassterm} since gauge invariance forbids such a term in the component expansion. A Standard Model fermion mass has to include a Higgs insertion (or appropriate composites neglected in this analysis). The operator of interest is of the form
\begin{equation}
 \frac{1}{\widetilde{M}^{2+n}}C_L C_R H M f(\Phi_{\rm hidden})
\end{equation}
where $f(\Phi_{\rm hidden})$ depends on further hidden sector fields that get vevs and has dimension $n$. 
Dimensional analysis requires the suppression by a factor  which  has mass dimensions ${2+n}$, $\widetilde{M}^{2+n}.$
This effective interaction arises from a diagram of type shown in Figure~\ref{diagrscheme}.

\begin{center}
\includegraphics[width=0.5\textwidth]{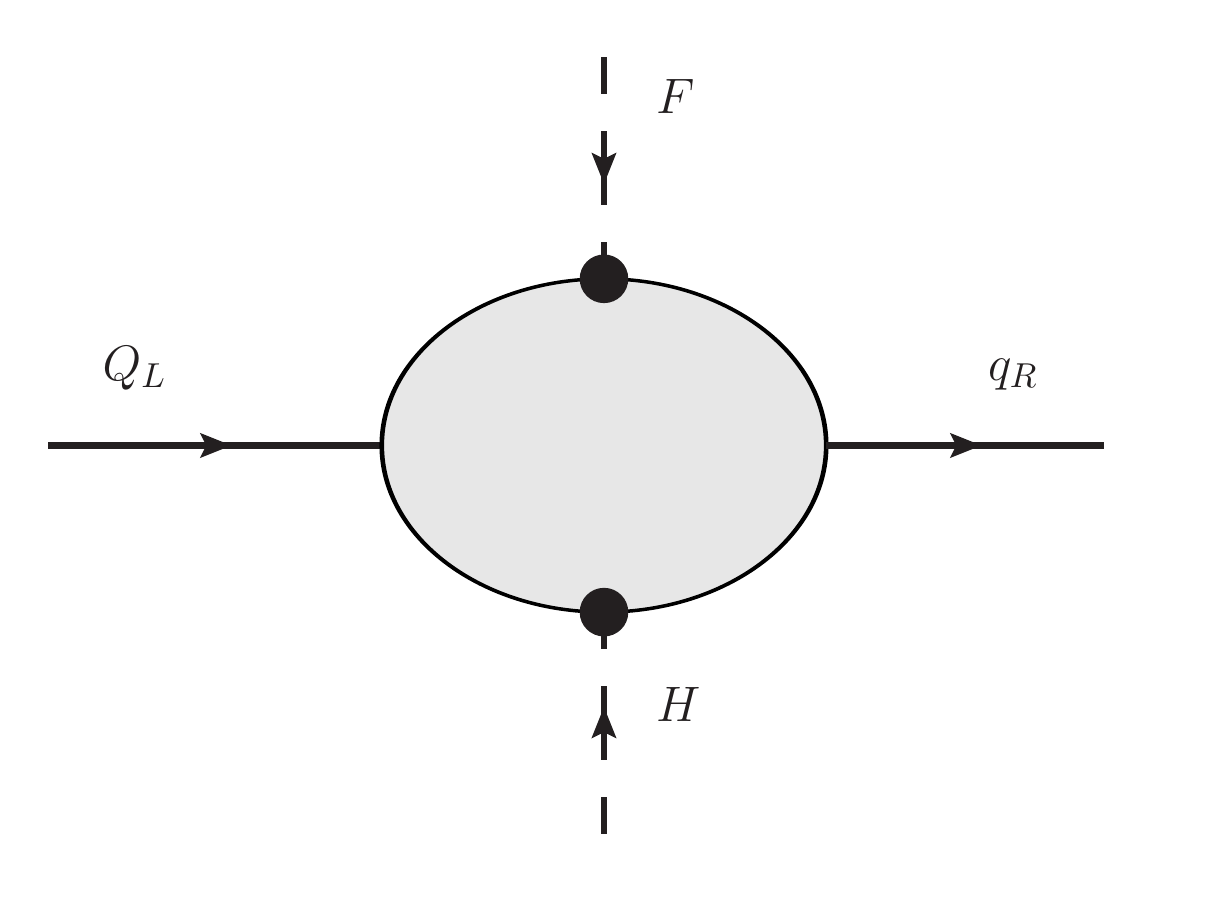}
\captionof{figure}{\footnotesize{Schematic diagram for radiative corrections to quark masses. The diagram needs to include one Higgs insertion and one SUSY breaking F-term insertion.}\label{diagrscheme}}
\end{center}
The contribution to fermion masses from the effective operator can be estimated to be
\begin{equation}
\label{newesti}
 \left.\frac{1}{\tilde{M}^{2+n}}C_L C_R H M f(\Phi_{\rm hidden})\right|_{\theta\theta\bar{\theta}\bar{\theta}}\sim {M_{\rm weak}}C_L C_R \frac{F_M}{\widetilde{M}^{2}}\, ,
\end{equation}
where in the last step we specialised to the dominant case of $n=0.$  There are two important mass scales scales 
 which are relevant for determining the suppression factor ${\widetilde{M}^{2}}$. The  first is the ultraviolet scale $M_{\rm{UV}}$, suppressing the mediation of SUSY breaking to the Standard Model sector by this factor. The second is the mass of the particles which have flavour changing interactions and run in the loops. 
 Motivated by absence of flavour changing processes at low energies, we begin our discussion by taking the the masses of these
particles to be the same as the ultraviolet scale\footnote{We shall take the flavour violating
couplings for these interactions at the high scale to be of order one.}  $M_{\rm{UV}}$. In this case there
is a single scale which determines $\widetilde{M}$, $\widetilde{M} \sim M_{UV}$. Thus one has a radiative
contribution to fermion masses
\begin{equation}
\label{finalest}
 \left.\frac{1}{M_{\rm UV}^{2}}C_L C_R H M f(\Phi_{\rm hidden})\right|_{\theta\theta\bar{\theta}\bar{\theta}}\sim \frac{M_{\rm weak}}{M_{\rm UV}}C_L C_R \frac{F_M}{M_{\rm UV}}\, ,
\end{equation}
 Thus the fermion masses are suppressed compared to the scalar masses by a factor of $M_{\rm higgs}/M_{\rm UV}.$ Assuming TeV scale supersymmetry, we then can estimate the suppression compared to the scalar masses and find unrealistic fermion masses for the lightest generation.
\begin{equation}
 m_{\rm fermion}\sim m_{\rm scalar} \frac{M_{\rm higgs}}{M_{\rm UV}}\sim 1\, {\rm TeV} \frac{10^3}{10^{18}}\sim 10^{-3} {\rm eV}\, . \label{msugraestimate}
\end{equation}
This summarises the results of ref.~\cite{Ibanez:1982xg}.

However in the above estimate we assume that the gravitino mass is of the same order as the scalar masses. The contribution to the fermion masses depends on the scale of the gravitino mass which can be parametrically larger than the scalar masses, if for example large no-scale cancellations take place for the scalar masses~\cite{0610129,0906.3297}. Furthermore, $M_{\rm{UV}}$ can be parametrically smaller than $M_{\rm{Planck}}$ in string constructions. Together these facts can provide an interesting loop-hole to the above argument, we now would like to explore this possibility in detail in the LVS setting.

\subsection{SUSY Breaking in the LVS}

As reviewed earlier, the LVS allows for moduli stabilisation for both K\"ahler and complex structure moduli. At the LVS minimum, supersymmetry is broken by F-terms associated to the K\"ahler moduli. Due to various no-scale cancelations, the spectrum of supersymmetric soft-masses is non-generic from a CMSSM point of view. The spectrum has been extensively studied for constructions with the Standard Model on the same cycle as the non-perturbative effects~\cite{0505076, 0610129} (scenario A) and for the Standard Model constructions separated from the non-perturbative effects~\cite{0906.3297} (scenario B). Details of the calculation can be found in those papers and we restrict ourselves here to summarising the hierarchical suppression of the soft-masses.
\begin{center}
\begin{tabular}{|c|c|}
\hline
$m_{3/2}$ & $\frac{M_P}{{\cal V}}$ \\ \hline
$M_{\rm gaugino}$ & $\frac{m_{3/2}}{\log(\cal  V)}$\\ \hline
$m_{\rm scalar}$ & $\frac{m_{3/2}}{\log(\cal  V)}$\\  \hline
$M_{\rm UV}$ & $M_{\rm string}$\\ \hline
$M_{\rm string}$ & $\frac{M_P}{\sqrt{{\cal V}}}$\\ \hline
${\cal V}$ & $10^{15}$\\ \hline
\end{tabular}
\captionof{table}{\footnotesize{Summary of hierarchical suppression of important masses for supersymmetry breaking as studied in scenario 1}.} \label{softmassesoriginal}
\end{center}
The scales for scenario A are summarised in Table~\ref{softmassesoriginal}. The cut-off scale, $M_{\rm UV}$, is the string scale since both gaugino and scalar masses are induced from the F-term of the local cycle where the Standard Model construction is placed. 

Shortly after, it was realised that the non-perturbative effects (gaugino condensation or E3-branes) responsible for moduli stabilisation cannot be present on the Standard Model cycle~\cite{0711.3389}. In this scenario SUSY is broken on a hidden cycle. As discussed in~\cite{0906.3297}, there are two options for the soft-masses depending on possible further no-scale cancelations due to uncertainties in the matter metrics. The scales of soft-masses are summarised in Table~\ref{softmassesnew}. In this scenario gaugino masses are induced from the dilaton F-term and scalar masses are induced from F-terms associated to cycles different from the Standard Model cycle, thus $M_{\rm UV}=M_{\rm Planck}.$

\begin{center}
\begin{tabular}{|c|c|}
\hline
$m_{3/2}$ & $\frac{M_P}{{\cal V}}$ \\ \hline
$M_{\rm gaugino}$ & $\frac{m_{3/2}}{\cal V}$\\ \hline
$m_{\rm scalar}$ & $\frac{m_{3/2}}{\sqrt{\cal V}}$ or $\frac{m_{3/2}}{\cal V}$\\ \hline
$M_{\rm UV}$ & $M_{\rm Planck}$\\ \hline
$M_{\rm string}$ & $\frac{M_P}{\sqrt{{\cal V}}}$\\ \hline
${\cal V}$ & $10^{6-7}$\\ \hline
\end{tabular}
\captionof{table}{\footnotesize{Summary of hierarchical suppression of important masses for supersymmetry breaking as studied in~\cite{0906.3297}. The choice in the scalar masses depends on additional cancellations due to the structure of matter metrics (cf.~\cite{0906.3297} for more details).}\label{softmassesnew}}
\end{center}

\subsection{Radiative fermion masses in the LVS}

Let us now estimate the size for radiatively generated fermion  for both scenarios. In scenario A  the radiatively generated masses for the fermions are
\begin{equation}
 m_{\rm rad}\sim M_{\rm weak} \frac{F_M}{M_{\rm UV}^2}\sim M_{\rm weak} \frac{1}{\sqrt{\cal V}}\, .
\end{equation}
Note that the suppression is given in terms of the volume in string units and compared to the weak scale is suppressed by a factor of square root the volume. With a volume of order $10^{15},$ we obtain a mass of the order of $100\, {\rm eV}.$ While this is a significant improvement compared to the previous discussion (cf. Equation~\ref{msugraestimate}), this is still not in the experimentally observed range.

In scenario B one similarly obtains the following suppression
\begin{equation}
 m_{\rm rad}\sim M_{\rm weak} \frac{F_M}{M_{\rm UV}^2}\sim M_{\rm weak} \frac{1}{{\cal V}^{3/2}}\, .
\end{equation}
We obtain a higher suppression in terms of the volume but for TeV soft masses the volume is smaller and we estimate the radiative corrections to be of order $1\, {\rm eV}.$

In the above estimates we used the relevant F-term for soft-masses. We would like to note that the F-term of the large K\"ahler modulus, which is determining the gravitino mass, is hierarchically larger and would significantly improve the estimate in the second scenario
\begin{equation}
   m_{\rm rad} \sim M_{\rm weak} \frac{1}{\cal{V}}\, ,
\end{equation}
leading to radiative masses in the MeV scale. These estimates are a significant improvement over the generic gravity mediation
case and provide encouragement for detailed loop calculations  and studies of RG evolution.\footnote{We mention in passing that a new scenario for soft breaking terms
has recently emerged in the context of LVS~\cite{cjp, Choi}. In this scenario a crude estimate of the radiatively generated masses gives a result similar to~\cite{Ibanez:1982xg}.}

\subsection{Explicit Mechanisms}
\label{expli}
Our discussion so far has focused on higher dimensional operators leading to radiative fermion masses. We now would like to give a realisation in terms of Feynman diagrams in the context of branes at singularities. As discussed before, there are two types of contributions which can lead to the desired fermion masses for the lightest generation:
\begin{enumerate}
\item Diagrams involving additional Higgs fields that allow for mixing among flavours.
\item Diagrams involving gaugino contributions that arise from non-flavour diagonal soft masses.
\end{enumerate}
Examples of both types are shown in Figure~\ref{diagram2}. Having multiple Higgs fields in the Standard Model arising from del Pezzo singularities, we start by estimating contributions from additional Higgses and then comment on potential contributions from flavour violating soft masses.
\begin{center}
\begin{tabular}{c c}
\includegraphics[width=0.45\textwidth]{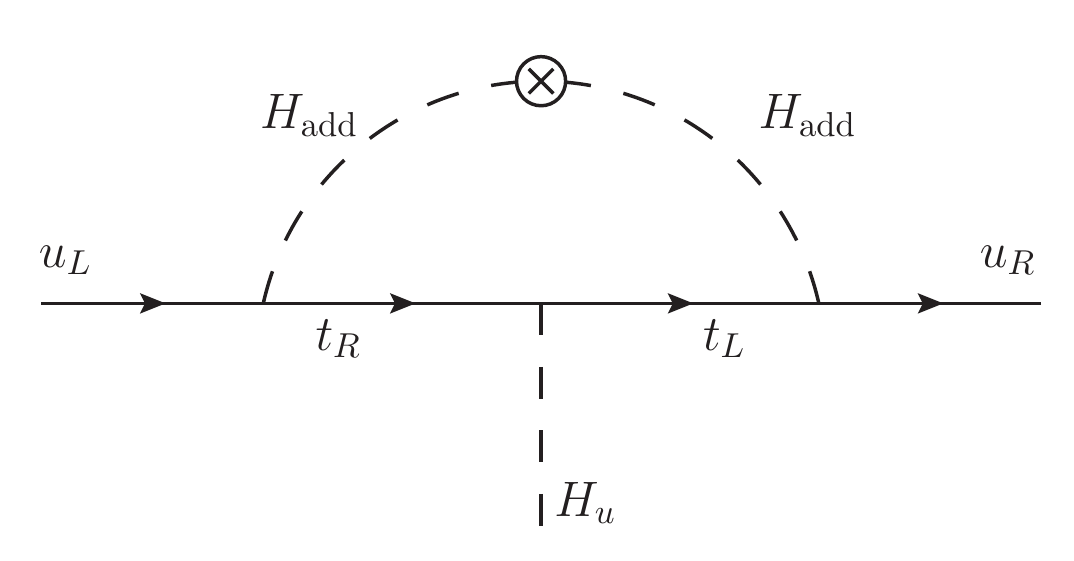} & \includegraphics[width=0.5\textwidth]{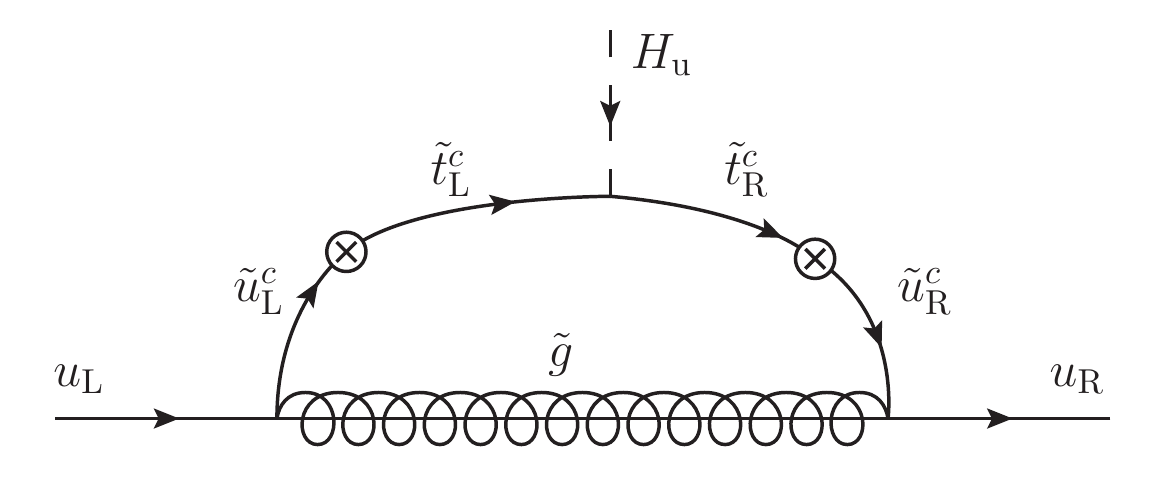}
\end{tabular}
\captionof{figure}{\footnotesize{{\bf Left:} 1-loop contribution to the up-mass involving additional Higgs fields. {\bf Right:} 1-loop induced up-quark mass involving gluinos and using non-diagonal entries in the squark mass matrix~\cite{Hall:1985dx} .}\label{diagram2}}
\end{center}
Such diagrams are typical for fermion masses in models with additional Higgses (for a review see~\cite{0910.2948}). The size of this diagram is suppressed by the inverse power of the largest mass in the loop. In our case, this is the additional Higgs and we then find the contribution to be given by
\begin{equation}
\label{addi}
m_{\rm up}\sim m_t \frac{F_{\rm SUSY}}{m_{H_{\rm add}}^2}\, .
\end{equation}
The SUSY insertion is at the additional Higgs and is needed here to give a non-vanishing effect. This SUSY insertion is a scalar soft mass term, hence smaller than the gravitino mass in the second scenario. To obtain the desired suppression, the additional Higgs mass has to be sufficiently larger than the SUSY insertion. In fact taking the scalar masses to scale as $1/{\cal V}^2$ - implying a volume ${\cal V}\sim 10^6$, the Higgs mass has to be at  the string scale for a suppression of the desired order of magnitude. Such a high scale Higgs mass is not only well above the experimental limits but also very interesting in the context of gauge coupling unification, which seem to require independently that these Higgs fields shall have a large mass and shall not participate in the low-energy running.

The radiative mass arising from the second diagram in Figure~\ref{diagram2} depends crucially on the off-diagonal contribution in the sfermion mass matrix. Defining $\epsilon $ as the ratio between the diagonal and off-diagonal entries in the sfermion mass matrix~\cite{Hall:1985dx}, one finds
\begin{equation}
\label{mixing}
 m_{\rm up}\sim m_{\tilde{t}}\, \epsilon^2\, .
\end{equation}
Constraints from neutral meson mixing provide strong bounds, $\epsilon < 10^{-3} \frac{M_{\rm susy}}{500 {\rm{GeV} }}$~\cite{luty}, posing a challenge to obtain masses in the phenomenologically interesting range.

Notice that both (\ref{addi}) and (\ref{mixing}) would agree with (\ref{finalest}) if $\tilde{M}$ were interpreted as the largest mass scale in the loop. 

\section{Conclusions}

In this paper we have studied radiative masses for the lightest generation of Standard Model fermions in local D-brane models. For del Pezzo singularities, we identified the global symmetries that forbid the mass at tree level and the effective operators that can break these global symmetries.

We estimated the effect of radiative corrections in generating non-zero values for the fermion masses. We found that for models based on the LVS these contributions are substantially larger than in generic gravity mediated scenarios and can be in the phenomenologically attractive range. This motivates detailed study of explicit processes and RG running. We have discussed scenarios involving additional Higgs fields and flavour changing effects in the sfermion mass matrix which can give fermion masses within the realistic range.

Other mechanisms for fermion mass generation can be relevant. One can consider non-trivial B-flux  threading the cycles of the singularity~\cite{Wijnholt:2005mp}, a similar mechanism has played an important role in F-theory models~\cite{Heckman:2010pv, 0910.0477}. Non-perturbative effects could become relevant with a Standard Model brane construction in the geometric regime.

\section*{Acknowledgments}

We would like to thank Kaladi Babu, Joe Conlon, Gia Dvali, Luis Ib\'a\~nez, Noppadol Mekareeya, Stuart Raby, Graham Ross and James Wells for useful discussions. CB, SK and AM thank the Abdus Salam International Centre for Theoretical Physics for its kind hospitality while part of this work was done, and Eyjafjallajokull for helping to provide us with some unexpected but undivided time. SK would like to thank the Abdus Salam International Centre for Theoretical Physics for financial support. The work of AM was supported by the EU through the Seventh Framework Programme and University of Cambridge. CB's research was supported in part by funds from the Natural Sciences and Engineering Research Council (NSERC) of Canada. Research at the Perimeter Institute is supported in part by the Government of Canada through Industry Canada, and by the Province of Ontario through the Ministry of Research and Information (MRI).
\newpage
\appendix
\section{The zero mass in $dP_2$}

  In this appendix we examine the $dP_2$ quiver and show that a mass term for the lightest up quark
is forbidden by the global symmetries. We start with the following superpotential in $dP_2$
\begin{equation}
W=\left(
\begin{array}{c}
 X_{43} \\
 Y_{23} \\
 Z_{23}
\end{array}
\right)\left(
\begin{array}{ccc}
 0 & Z_{14} & -Y_{64} \\
 -Z_{14} \Phi _{61} \frac{\Psi _{42}}{\Lambda^2} & 0 & X_{12} \frac{\Phi _{61}}{\Lambda} \\
 Y_{64} \frac{\Psi _{42}}{\Lambda} & -X_{12} & 0
\end{array}
\right)\left(
\begin{array}{c}
 X_{36} \\
 Y_{31} \\
 Z_{36}
\end{array}
\right),
\end{equation}
where $X_{43}, Y_{23},$ and $Z_{23}$ denote left-handed quarks and respectively $ X_{36}, Y_{31},$ and $ Z_{36}$ right handed quarks. In addition to the anomalous $U(1)$ symmetry of the individual $U(n_i)$ gauge group, the superpotential is characterised by a global $SU(2)\times U(1)\times U(1)_R$ symmetry. The charges of the individual states under these global symmetries are summarised in Table~\ref{tab:chargesdp2}.

In analogy to the discussion of $dP_1$ we can go to the mass eigenbasis using the following unitary matrices $A$ and $B:$
\begin{eqnarray}
A&=&\left(
\begin{array}{ccc}
 a_1 X_{12} \Phi _{61} & a_{12} & a_{13} \\
 a_1 Y_{64} & a_{22}  ( \Phi _{61}X_{12})^* Y_{64} & a_{23} ( \Phi _{61} X_{12})^* Y_{64} \\
 a_1 Z_{14} \Phi _{61} & a_{32} (X_{12})^* Z_{14} & a_{33}(X_{12})^* Z_{14}
\end{array}
\right)\\
B&=& \left(
\begin{array}{ccc}
 b_1 X_{12} & b_{12}(\Psi _{42})^* & -b_{13}(\Psi _{42})^* \\
 b_1 Y_{64} \Psi _{42} & b_{22} (X_{12})^*Y_{64} &  b_{23}(X_{12})^* Y_{64} \\
 b_1 Z_{14} \Psi _{42} & b_{32}(X_{12})^*  Z_{14} & b_{33} (X_{12})^*  Z_{14}
\end{array}
\right),
\end{eqnarray}
where $a_{ij}$ and $b_{ij}$ absorb gauge invariant factors (omitted for clarity) and gauge invariant normalisation factors. In this basis we can write the superpotential as
\begin{eqnarray}
\nonumber W&=&\left(\begin{array}{c}
X_{43}(\Phi_{61}X_{12})^*+(Y_{64})^*Y_{23}+(\Phi_{61}Z_{14})^*Z_{23}\\
(Z_{14})^*X_{12}Z_{23}+(Y_{64})^*\Phi_{61}X_{12}Y_{23}+X_{43}\\
(Z_{14})^*X_{12}Z_{23}+(Y_{64})^*\Phi_{61}X_{12}Y_{23}+X_{43}
\end{array}\right).
\left(\begin{array}{c c c}
0 & 0 & 0\\
0 & Y_{64}(X_{12})^*Z_{14}& 0\\
0 & 0 & Y_{64}(X_{12})^*Z_{14}
\end{array}
\right).\\ &&
\left(\begin{array}{c}
(X_{12})^* X_{36}+(Y_{64} \Psi _{42})^* Y_{31}+(Z_{14} \Psi _{42})^* Z_{36}\\
Y_{31}X_{12}(Y_{64})^*+X_{36}\Psi_{42}+Z_{36}(Z_{14})^*X_{12}\\
Y_{31}X_{12}(Y_{64})^*+X_{36}\Psi_{42}+Z_{36}(Z_{14})^*X_{12}
\end{array}\right),
\end{eqnarray}
where we have omitted the gauge invariant normalization factors. We can see that a mass term for the lightest generation requires a term of the form $A_{26}B_{26} C_{14}.$ Such a term involving only $D3D3$ states cannot lead to a gauge invariant operator with R-charge $2,$ hence  the global symmetries of $dP_2$ forbid a mass for the lightest generation. In $dP_2$ we see that the anomalous $U(1)$ symmetries do not forbid the mass for the lightest generation, but the $U(1)_R$ symmetry plays the crucial role.
\begin{center}
\begin{tabular}{c |c |c |c |c |c |c |c| c}
Field & $SU(2)$ & $U(1)$ & $U(1)_R$ & $n_1$ & $n_2$ & $n_3$ & $n_4$ & $n_6$\\ \hline \hline
$X_{43}$ & 1& 2& $\frac{6}{7}$ &0 &0 & -1&1 &0 \\
$Y_{23}$ & 2& -1&$\frac{4}{7}$ & 0& 1&-1 &0 &0 \\
$Z_{23}$ & 2& -1&$\frac{4}{7}$ &0 & 1&-1 &0 &0 \\
\hline
$X_{36}$ &2 & -1&$\frac{4}{7}$ & 0& 0&1 &0 &-1 \\
$Y_{31}$ & 1& 2&$\frac{6}{7}$ &-1 &0 &1 &0 &0 \\
$Z_{36}$ &2 &-1 & $\frac{4}{7}$& 0& 0& 1& 0& -1\\
\hline
$Y_{64}$ & 2&-1 &$\frac{4}{7}$ &0 &0 &0 &-1 &1 \\
$X_{12}$ & 2&-1 &$\frac{4}{7}$ & 1& -1 &0 &0 &0 \\
$Z_{14}$ & 1&-4 & $\frac{2}{7}$&1 &0 &0 &-1 &0 \\
\hline
$\Phi_{61}$ &2 &3 & $\frac{2}{7}$&-1 &0 &0 &0 &1 \\
$\Psi_{42}$ & 2&3 & $\frac{2}{7}$& 0&-1 &0 &1 &0 \\
\hline
$Q_{Lm}^1$ & 0&0 &$\frac{8}{7}$ & 0&1 &-1 &1 &-1 \\
$Q_{Lm}^2$ & 1&2 & $\frac{6}{7}$& 0&0 &-1 &1 &0 \\
$Q_{Lm}^3$ & 1&2 & $\frac{6}{7}$&0 &0 &-1 &1 &0 \\
\hline
$Q_{Rm}^1$ &0 &0 &$\frac{8}{7}$ &-1 &1 & 1&0 &-1 \\
$Q_{Rm}^2$ &1 &2 & $\frac{6}{7}$ & 0&-1 &1 &1 &-1 \\
$Q_{Rm}^3$ & 1& 2& $\frac{6}{7}$& 0& -1& 1& 1& -1 \\
\hline
\end{tabular}
\captionof{table}{{\footnotesize Summary of the charges under the global symmetry for the fields in $dP_2.$ The last six fields denote the quarks in the mass eigenstate basis.} \label{tab:chargesdp2}}

\end{center}

\bibliography{rad}
\bibliographystyle{JHEP}
\end{document}